# Ultra-low-power orbital-controlled magnetization switching using a ferromagnetic oxide interface

Le Duc Anh,[1,2,*] Takashi Yamashita,[1] Hiroki Yamasaki,[1] Daisei Araki,[1] Munetoshi Seki,[1,3] Hitoshi Tabata,[1,3] Masaaki Tanaka[1,3,**] and Shinobu Ohya[1,2,3,***]

[1]*Department of Electrical Engineering and Information Systems, The University of Tokyo, 7-3-1 Hongo, Bunkyo-ku, Tokyo 113-8656, Japan*
[2]*Institute of Engineering Innovation, Graduate School of Engineering, The University of Tokyo, 7-3-1 Hongo, Bunkyo-ku, Tokyo 113-8656, Japan*
[3]*Center for Spintronics Research Network (CSRN), The University of Tokyo, 7-3-1 Hongo, Bunkyo-ku, Tokyo 113-8656, Japan*

Email: *anh@cryst.t.u-tokyo.ac.jp,
**masaaki@ee.t.u-tokyo.ac.jp,
***ohya@cryst.t.u-tokyo.ac.jp

**A major challenge in spin-based electronics is reducing power consumption for magnetisation switching of ferromagnets, which is being implemented by injecting a large spin-polarized current. The alternative approach is to control the magnetic anisotropy (MA) of the ferromagnet by an electric field. However, the voltage-induced MA is too weak to deterministically switch the magnetisation without an assisting magnetic field, and the strategy towards this goal remains elusive. Here, we demonstrate a new scheme of orbital-controlled magnetisation switching (OCMS): A sharp change in the MA is induced when the Fermi level is moved between energy bands with different orbital symmetries. Using a ferromagnetic oxide interface, we show that OCMS can be used to achieve a deterministic and magnetic-field-free 90°-magnetisation switching solely by applying an extremely small electric field of 0.05 V/nm with a negligibly small current density of ~ $10^{-2}$ A/cm². Our results highlight the huge potential of band engineering in ferromagnetic materials for efficient magnetisation control.**

In present spin-based electronic devices such as magnetic tunnel junctions (MTJs), magnetization switching is being implemented by either spin transfer torque (STT) [1-3] or spin-orbit torque (SOT) [4,5]. These current-driven switching methods require a very large current density, typically in the order of $10^6$–$10^7$ A/cm$^2$ for STT and $10^5$–$10^6$ A/cm$^2$ for SOT, which causes high-power consumption, and thus degrades the device endurance and scalability. Therefore, there is an intensified interest in new electric field-driven schemes for magnetization switching, whose key ingredient is to manipulate the magnetic anisotropy (MA) of the ferromagnetic thin films using a bias voltage [6–16]. However, in ferromagnetic metals [11-14], the electron charging effect by an electric field is effective only in monolayer-thin films, and the modification is at most 10% of the total interfacial MA energy for a typical electric field of 1 V/nm, as also expected from density functional theory modelling [17]. Because of this small change in the MA energy, the demonstration of electric field-driven magnetization switching always requires an external magnetic field or a well-calibrated sequence of ultrafast electrical pulses [13], both of which are impractical for device applications. The magnetoelectric effect in multiferroic materials such as BiFeO$_3$ has shown to be promising for reducing the power consumption of magnetization switching by an order of magnitude [15]; however, in those cases, the choice of materials is limited and the very low device endurance is a problem.

In this Letter, we represent that a highly effective magnetization switching can be implemented by moving the Fermi level ($E_F$) between energy bands with different-symmetry orbitals, which is induced by a strong change of the MA via spin-orbit interactions [Fig. 1(a)]. We call this new scheme *orbital-controlled magnetization switching* (OCMS). Using an MTJ composed of La$_{0.67}$Sr$_{0.33}$MnO$_3$ (LSMO)/SrTiO$_3$ (STO)/LSMO as a model system, we demonstrate a *deterministic magnetic-field-free* sharp 90°-switching of the magnetization induced solely by a small change in the electric field (~0.05 V/nm) applied on the tunnel barrier and a negligibly small current density (~ $10^{-2}$ A/cm$^2$). These results also suggest that OCMS can possibly be applied to a wide range of materials that have only moderate spin-orbit interaction like LSMO if the band structure is appropriately designed.

For demonstrating OCMS, LSMO [18–20] is an ideal ferromagnetic material thanks to its band structure: Immediately (~0.2 eV) below the $E_F$ in LSMO at the LSMO/STO interface, there are the bottom band edge of the up-spin $e_g$ orbitals and the top band edge of the $t_{2g}$ orbitals [21,22]. We study an MTJ structure consisting of, from the top surface, LSMO [18 unit cell (u.c.) = 7.0 nm]/STO (10 u.c. = 3.9 nm)/LSMO (40 u.c. = 15.6 nm), which was etched into a circular mesa with a diameter of 800 μm [see Fig. 1(b) and Supplementary Material (SM) [23]]. As illustrated in Fig. 1(c), when applying a positive bias voltage $V$, for example, carrier depletion occurs at the top interface between LSMO and STO (left-side interface), which moves the quasi $E_F$ downward from the $e_g$ band to the $t_{2g}$ band and induces the sharp change in the MA of the top LSMO layer. For detecting this change in the MA when changing $V$, we measured the characteristics of tunneling magnetoresistance (TMR) with a magnetic field **H** applied along various in-plane directions (Fig. 2). Here, the TMR ratio is defined as [$R(\mathbf{H})$– $R(0)$]/$R(\mathbf{H})$, where $R(\mathbf{H})$ represents the tunnel resistance under **H**, and $\theta_H$, $\theta_{Mt}$, and $\theta_{Mb}$ are the in-plane angles of **H**, the magnetization vector of the top LSMO (**M**$_t$) layer, and that of the bottom LSMO (**M**$_b$) layer, respectively, from the [100] axis in the counter-clockwise direction as shown in Fig. 1(b). In Figs. 2(a) and 2(b), at small $V$ (15 mV), large and clear TMR switching occurs when **H** is along the [$\bar{1}$10] axis (a), while it is suppressed when **H** is along the [110] axis (b). However, as |$V$| is increased, the TMR

ratios measured with **H**//[110] and **H**//[$\bar{1}$10] become nearly identical. This anomalous feature can be more clearly seen in the $V$ dependence of the TMR for both **H** directions in Fig. 2(c). The color-coded polar plots of TMR($H$,$\theta_H$) in Fig. 2(d) reveal a sharp transition from a dominantly two-fold symmetry, with an easy axis lying along the [$\bar{1}$10] direction (see the red regions), to an almost four-fold symmetry along the <110> axes with increasing $V$ (see the emerging yellow regions along the [110] axis). This result means that the MA of the top LSMO layer is indeed changed from two-fold to four-fold by increasing $V$.

Using this sharp change in the MA, we can realize a deterministic magnetization switching of the top LSMO layer *without* an assistance of an external magnetic field. When $V$ is changed from 200 mV to 15 mV, the easy axis along [110] at $V$ = 200 mV becomes a hard axis at $V$ = 15 mV, while the [1$\bar{1}$0] axis is always an easy axis. Thus, the magnetization initially pointed toward [110] at $V$ = 200 mV is rotated toward [1$\bar{1}$0] when $V$ is changed to 15 mV. Then, after $V$ is returned to 200 mV from 15 mV, the magnetization remains in that position ([1$\bar{1}$0]). Meanwhile, when the magnetization is initially pointed to [$\bar{1}$10] at $V$ = 200 mV, it remains fixed during the change in $V$.

To verify this scenario, after applying **H** = 10 kOe along [110] or [$\bar{1}$10] to align **M**$_t$ and **M**$_b$ with $V$ = 200 mV, we measured the tunneling resistance $R$ at zero magnetic field by varying $V$ in the following sequence: 200 mV → 15 mV → 200 mV. In Figs. 3(a) and 3(b), the dotted horizontal lines $R_{P200}$ ($R_{P15}$) and $R_{AP200}$ ($R_{AP15}$) express the $R$ values at **H** = 0 when **M**$_t$ and **M**$_b$ are initialized in the parallel and antiparallel configurations at $V$ = 200 mV (15 mV), respectively [23]. When **H** // [110] [Fig. 3(a)], one can see that the $R$ value after the bias sequence does not return to $R_{P200}$, which indicates that **M**$_t$ is rotated during the bias sequence. We estimate the relative angle $\Delta\theta$ of the magnetizations between the LSMO layers after the bias sequence, using the relation (see SM [23])

$$\cos \Delta\theta = \frac{2R_{P200}R_{AP200} - RR_{AP200} - RR_{P200}}{R(R_{AP200} - R_{P200})}. \tag{1}$$

The estimated $\Delta\theta$ value is 87.7°, which is in excellent agreement with the above expectation of the 90° switching of the magnetization. Furthermore, when **H** // [$\bar{1}$10] [Fig. 3(b)], the MTJ consistently remained in the parallel magnetization configuration (*i.e.*, $R = R_{P15}$ or $R_{P200}$) as expected. The magnetization switching occurs with the electric field of only 0.05 V/nm (= 0.2 V/3.9 nm), which is two orders of magnitude smaller than that needed for voltage-controlled MA (~1 V/nm) in previous reports [10–14]. The current density during this bias sequence is negligibly small: $1.3\times10^{-2}$ A/cm$^2$ (at $V = 200$ mV) and $5.9\times10^{-4}$ A/cm$^2$ (at $V = 15$ mV).

As shown below, this unusually large change in the MA is related to the change in the orbital at the quasi $E_F$ in the top LSMO layer. This result can be understood from the correlation between the changes in the MA fields and in the density of states (DOS) when changing the magnetization direction. The derived MA fields of the LSMO layers, which we determined from the anisotropic TMR shown in Fig. 2(d) (see SM [23]), are summarized in Table 1; when increasing $V$ from 15 mV to 200 mV, the biaxial anisotropy field $H_{4<110>}$ increases by 40% (100 Oe), while the uniaxial anisotropy fields $H_{2[110]}$ and $H_{2[100]}$ rapidly decay and decrease by 85% (110 Oe) and 69% (45 Oe), respectively. This finding is consistent with the transition of the symmetry of the MA from two-fold to four-fold as shown in Fig. 2(d). Using these MA fields, we calculated the magnetostatic energy $E_{mag}$ [Fig. 3(c)], which well reproduces the result of the magnetization switching shown in Figs. 3(a) and (b); the local minimum of $E_{mag}$ at [110] when $V = 200$ mV (filled pink circle) becomes unstable when decreasing $V$ to 15 mV (dotted pink circle), while $E_{mag}$ is always minimum at [$\bar{1}$10] (filled grey circle). This result also confirms the validity of the derived MA fields.

The change in the DOS with the magnetization direction is obtained by measuring $dI/dV$ with a strong magnetic field of 1 T applied in various directions in the film plane, as shown in the color mapping of $\Delta(dI/dV)$ as a function of $V$ and $\theta_H$ (= $\theta_M$ at $\mathbf{H}$ = 1 T ) in Fig. 4(a) (see SM [23]). A clear change of $\sim \pm1.5\%$ in $dI/dV$ with two-fold symmetry was induced by rotating $\mathbf{H}$. This result indicates that the DOSs of the top and bottom LSMO layers change via the SOI when rotating the magnetization direction. The most important feature is that the direction $\theta_{DOS-MAX}$ of $\mathbf{H}$, the direction at which the DOS reaches its maximum, rotates by 90° when $V$ is varied through $V_p$ = 0.06 V–0.095 V and $V_n$ = (–0.15 V) – (–0.13 V), which means that the orbital symmetry at the $E_F$ is changed with $V$ [21]. We decomposed the oscillation component of the $dI/dV$–$\theta_M$ curves into a four-fold component $C_{4<110>}$ along <110> and two-fold components $C_{2[100]}$ and $C_{2[110]}$ along [100] and [110], respectively. (For the derivation of the components $C_{4<110>}$, $C_{2[100]}$, and $C_{2[110]}$, see SM [23]) As shown in Fig. 4(b), the absolute values of the two-fold symmetry MA fields ($H_{2[110]}$ and $H_{2[100]}$) and those of the symmetry DOS components (*i.e.* $C_{2[110]}$ and $C_{2[100]}$) show very similar dependences on $V$. These findings indicate that the large MA change is induced by the change in the orbital symmetry at the $E_F$. A similar strong correlation between the MA and DOS symmetry has also been reported for ferromagnetic semiconductor GaMnAs [24], suggesting the universality of this phenomena for a wide variety of materials. Our result highlights the huge potential of using band engineering techniques for manipulating the magnetization.

In this work, as the bias voltage increases, the two-fold MA disappears and is replaced with a four-fold one at the LSMO interface, which enables the one-way rotation of the magnetization from [110] to [1$\bar{1}$0]. For a reversible control of the magnetization direction, the direction of the two-fold or four-fold easy magnetization axis should be changed with the bias voltage [for example in Fig. 3(c), the two-fold easy magnetization axis at $V = 15$ mV should be switched from $[1\bar{1}0]$ to [110] at a different bias voltage]. Single-crystalline ferromagnetic quantum wells are promising candidates for the reversible magnetization rotation, in which itinerant $s,p$ electrons that are coupled with localized $d$ electrons make resonant levels. Using the ferromagnetic quantum well, alternating reversal of the sign of the DOS symmetry, which may lead to an alternating change of the easy magnetization axis, has been demonstrated when switching the on- and off-resonant states by changing the bias voltage [25].

Finally, we discuss the potential technological applicability of this OCMS method, which we have demonstrated at the LSMO interface only at low temperature (3.5 K) and in a large MTJ (800 μm in diameter) in this work. As shown in Fig. 3(c), the potential barrier between the [110] and $[1\bar{1}0]$ magnetization directions at $V = 200$ mV, which disappears at $V = 15$ mV, is estimated to be 53 Oe, which corresponds to an MA constant $K_b$ of 1.2 kJ/m³. This large $K_b$ indicates that the two states of [110] and $[1\bar{1}0]$ magnetization directions are thermally stable even at room temperature. At the present temperature (3.5 K), by keeping the thermal stability factor $\Delta$ (= $K_b v / k_B T$, where $v$ is the LSMO volume, $k_B$ is the Boltzmann constant and $T = 3.5$ K) of 60, which is usually required for 10 year data retention time [26], the size of our MTJ can be reduced as small as 21 nm in diameter. Furthermore, as the magnetization rotation is induced by the change in the MA energy, high-speed operation is in principle possible. Therefore, OCMS is technologically promising for state-of-the-art spin devices if we can appropriately design the band structure of ferromagnetic materials.

This work was partly supported by Grants-in-Aid for Scientific Research (No. 18H03860, 17H04922), the CREST Program (JPMJCR1777) of the Japan Science and Technology Agency, and the Spintronics Research Network of Japan (Spin-RNJ).

**References**

1. J. Slonczewski, Current-driven excitation of magnetic multilayers, J. Magn. Magn. Mater. **159**, L1 (1996).

2. L. Berger, Emission of spin waves by a magnetic multilayer traversed by a current. Phys. Rev. B **54**, 9353 (1996).

3. J. Z. Sun and D. C. Ralph, Magnetoresistance and spin-transfer torque in magnetic tunnel junctions. J. Magn. Magn. Mater. **320,** 1227 (2008).

4. I. M. Miron *et al*., Perpendicular switching of a single ferromagnetic layer induced by in-plane current injection. Nature **476**, 189 (2011).

5. L. Liu *et al*., Spin-torque switching with the giant spin Hall effect of tantalum. Science **336**, 555 (2012).

6. C-G. Duan *et al.,* Surface magnetoelectric effect in ferromagnetic metal films. Phys. Rev. Lett. **101**, 137201 (2008).

7. K. Nakamura *et al.,* Giant modification of the magnetocrystalline anisotropy in transition-metal monolayers by an external electric field. Phys. Rev. Lett. **102**, 187201 (2009).

8. J. Stohr, H. C. Siegmann, A. Kashuba, and S. J. Gamble, Magnetization switching without charge or spin currents. Appl. Phys. Lett. **94**, 072504 (2009).

9. N. N. Negulyaev, V. S. Stepanyuk, W. Hergert, and J. Kirschner, Electric field as a switching tool for magnetic states in atomic-scale nanostructures. Phys. Rev. Lett. **106**, 037202 (2011).

10. D. Chiba, M. Yamanouchi, F. Matsukura, and H. Ohno, Electrical manipulation of magnetization reversal in a ferromagnetic semiconductor. Science **301**, 943 (2003).

11. W-G. Wang, M. Li, S. Hageman, and C. L. Chien, Electric-field-assisted switching in magnetic tunnel junctions. Nat. Mater. **11,** 64 (2012).

12. T. Maruyama *et al.,* Large voltage-induced magnetic anisotropy change in a few atomic layers of iron. Nat. Nanotech. **4**, 158 (2009).

13. Y. Shiota *et al.,* Induction of coherent magnetization switching in a few atomic layers of FeCo using voltage pulses. Nat. Mater. **11**, 39 (2012).

14. S. Kanai, F. Matsukura, and H. Ohno, Electric-field-induced magnetization switching in CoFeB/MgO magnetic tunnel junctions. Appl. Phys. Lett. **108**, 192406 (2016).

15. J. T. Heron, J. L. Bosse, Q. He, Y. Gao, M. Trassin, L. Ye, J. D. Clarkson, C. Wang, J. Liu, S. Salahuddin, D. C. Ralph, D. G. Schlom, J. Iniguez, B. D. Huey, and R. Ramesh, Deterministic switching of ferromagnetism at room temperature using an electric field. Nature **516**, 370 (2014).

16. U. Bauer, L. Yao, A. J. Tan, P. Agrawal, S. Emori, H. L. Tuller, S. V. Dijken, and G.

S. D. Beach. Magneto-ionic control of interfacial magnetism. Nat. Mater. **14**, 174 (2015).

17. M. Tsujikawa, S. Haraguchi, T. Oda, Y. Miura, and M. Shirai, Giant magnetic tunneling effect in $Fe/Al_2O_3/Fe$ junction. A comparative ab initio study on electric-field dependence of magnetic anisotropy in MgO/Fe/Pt and MgO/Fe/Au films. J. Appl. Phys. **109**, 07C107 (2011).

18. A. Asamitsu, Y. Moritomo, Y. Tomioka, T. Arima, and Y. Tokura, A structural phase transition induced by an external magnetic field. Nature **373**, 407 (1995).

19. J-H. Park, E. Vescovo, H-J. Kim, C. Kwon, R. Ramesh, and T. Venkatesan, Direct evidence for a half-metallic ferromagnet. Nature **392**, 794 (1998).

20. M. Bowen *et al*., Spin-polarized tunneling spectroscopy in tunnel junctions with half-metallic electrodes. Phys. Rev. Lett. **95**, 137203 (2005).

21. L. D. Anh, N. Okamoto, M. Seki, H. Tabata, M. Tanaka, and S. Ohya, Hidden peculiar magnetic anisotropy at the interface in a ferromagnetic perovskite-oxide heterostructure. Sci. Rep. **7,** 8715 (2017).

22. J. D. Burton and E. Y. Tsymbal, Tunneling anisotropic magnetoresistance in a magnetic tunnel junction with half-metallic electrodes. Phys. Rev. B **93,** 24419 (2016).

23. See Supplemental Material at [URL will be inserted by publisher] for details on experimental and theoretical methods.

24. H. Saito, S. Yuasa, and K. Ando, Origin of the tunnel anisotropic magnetoresistance in $Ga_{1-x}Mn_xAs/ZnSe/Ga_{1-x}Mn_xAs$ magnetic tunnel junctions of II-VI/III-V heterostructures. Phys. Rev. Lett. **95**, 086604 (2005).

25. I. Muneta, T. Kanaki, S. Ohya, and M. Tanaka, Artificial control of the bias-voltage dependence of tunnelling-anisotropic magnetoresistance using quantization in a single-crystal ferromagnet. Nat. Commun. **8**, 15387 (2017).

26. A. D. Kent and D. C. Worledge, A new spin on magnetic memories. Nat. Nanotechnol. **10**, 187 (2015).

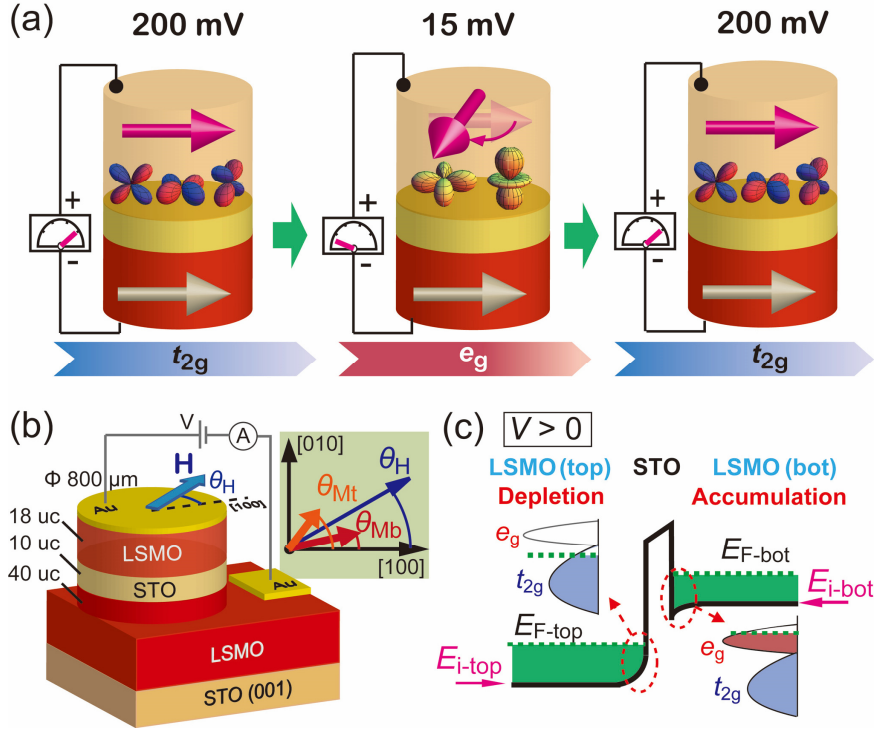

FIG. 1. (a) Deterministic and magnetic field-free 90°-magnetization switching, which is induced only by slightly changing the bias voltage $V$ applied to the MTJ. This phenomenon occurs with the change in the orbital symmetry between $e_g$ and $t_{2g}$ at the quasi-Fermi energy ($E_F$) of the top interface between LSMO [ferromagnetic (FM) layer] and STO (tunnel barrier). (b) Device structure and measurement configuration of the LSMO/STO/LSMO MTJ used in this study. (c) Band profile of the MTJ when $V$ is positive, where the green dotted lines represent the quasi-Fermi levels ($E_{F-top}$, $E_{F-bot}$) in the top and bottom LSMO layers. There are carrier depletion and accumulation at the top and bottom LSMO/STO interfaces, respectively. Thus, when increasing $V$ (>0), the band character of the carriers at $E_{F-top}$ changes from $e_g$ to $t_{2g}$ symmetry at the top LSMO/STO interface (left inset).

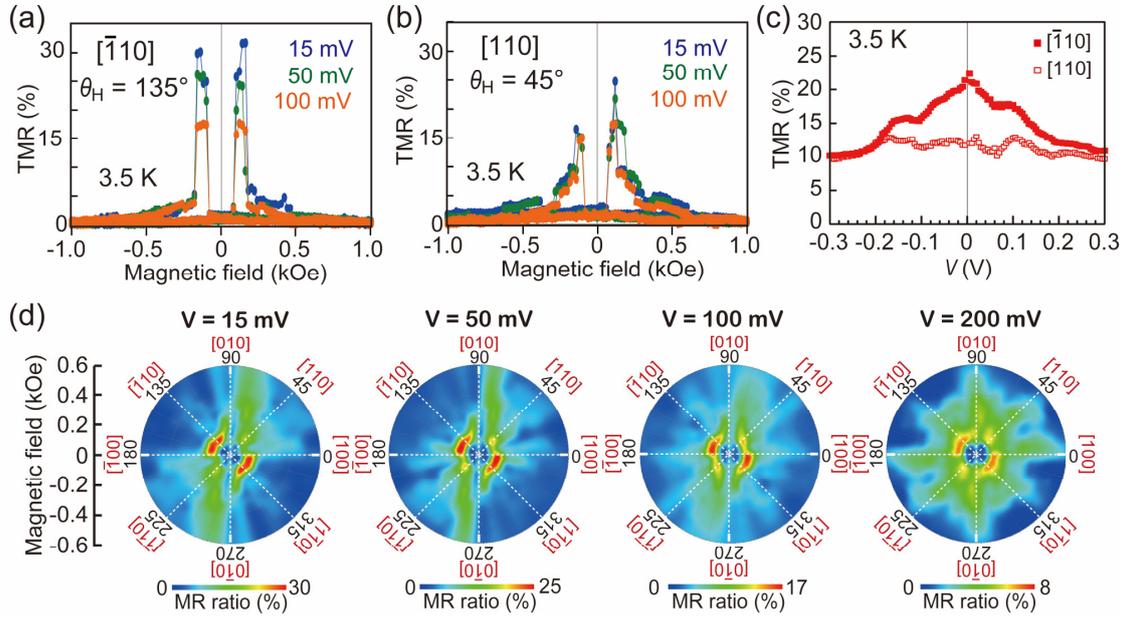

FIG. 2. (a)(b) **H** dependence of TMR (major loops) measured at $V$ = 15, 50, and 100 mV with **H** applied in the film plane along $[\bar{1}10]$ (a) and $[110]$ (b). (c) Bias dependence of the TMR ratio with **H** applied along $[\bar{1}10]$ and $[110]$. (d) Polar color-mapping plots of the magnetic-field-direction dependence of TMR under a bias voltage $V$ = 15, 50, 100, and 200 mV. All measurements were carried out at 3.5 K.

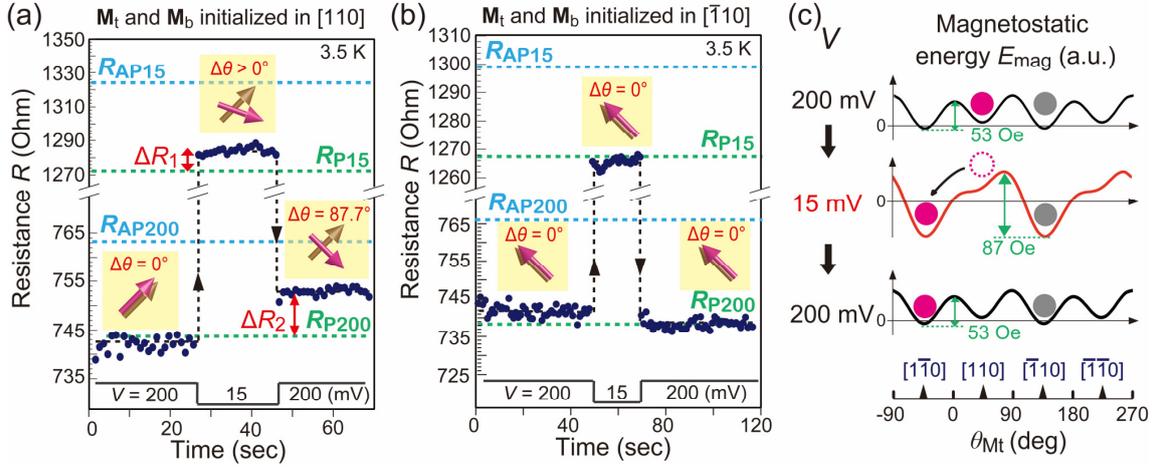

FIG. 3. (a)(b) Time evolution of the tunneling resistance $R$ of the MTJ (blue dots) measured at zero magnetic field with a bias sequence of $V = 200$ mV → 15 mV → 200 mV, after applying a strong **H** = 10 kOe along [110] (a) and [$\bar{1}$10] (b) to align the magnetization vectors of the top and bottom LSMO, **M**$_t$ and **M**$_b$. The pink and brown arrows in the yellow insets illustrate **M**$_t$ and **M**$_b$ in the film plane, respectively, which form an angle $\Delta\theta$ determined using Eq. (1). In (a), $R$ increased by $\Delta R_1$ from $R_{P15}$ and by $\Delta R_2$ from $R_{P200}$ after the bias sequence due to the switching of **M**$_t$. (c) Magnetostatic energy $E_{mag}$ as a function of the magnetization direction $\theta_{Mt}$ of the top LSMO at $V = 200$ mV and 15 mV, calculated using the estimated anisotropy fields in Table 1. The local minimum of $E_{mag}$ at [110] when $V = 200$ mV becomes unstable when $V$ is decreased to 15 mV, while that at [$\bar{1}$10] is always the minimum for all $V$. Thus, the magnetization initially pointed toward [110] (pink circle) at $V = 200$ mV rotates toward the [1$\bar{1}$0] direction at $V = 15$ mV, while the one initially pointed toward [$\bar{1}$10] (grey circle) at $V = 200$ mV stays fixed.

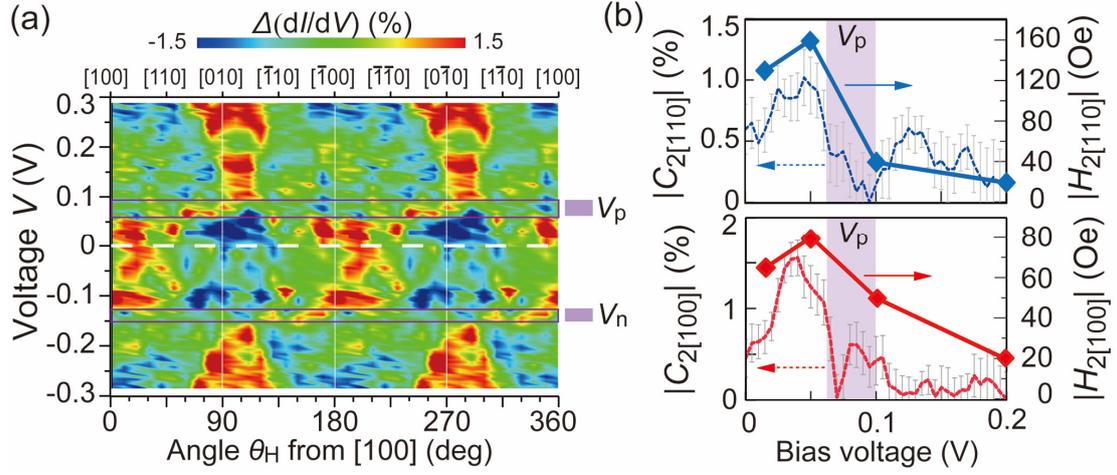

FIG. 4. (a) Color-mapping plot of $\Delta\left(\frac{\mathrm{d}I}{\mathrm{d}V}\right)$. The direction $\theta_{\mathrm{DOS\text{-}MAX}}$, along which $\mathrm{d}I/\mathrm{d}V$ reaches its maximum, rotates by 90° at $V = V_{\mathrm{p}}$ (= 0.6 V ─ 0.95 V) and $V_{\mathrm{n}}$ [= ( ─ 0.15 V) ─ (─ 0.13 V)] (purple bands). (b) Absolute values of the symmetry DOS components $C_{2[110]}$ and $C_{2[100]}$ deduced from the magnetization-direction dependence of the DOS (dotted lines with error bars), and those of the anisotropy fields $H_{2[110]}$ and $H_{2[100]}$ obtained from the TMR data at $V = 0.015$, 0.05, 0.1, and 0.2 V (closed rhombuses), as functions of the bias voltage $V$. All measurements were carried out at 3.5 K.

TABLE 1. Obtained parameters of the anisotropy fields: Biaxial anisotropy field $H_{4\langle110\rangle}$ along the <110> axes, uniaxial anisotropy field $H_{2[110]}$ along the [110] axis, uniaxial anisotropy field $H_{2[100]}$ along the [100] axis, the domain wall pinning energy $\varepsilon$, and the spin polarization $P$ of the top and bottom LSMO layers, obtained from the tunneling magnetoresistance data using the Stoner-Wohlfarth model.

| Bias voltage (V) | Top LSMO layer | | | | Bottom LSMO layer | | | | $P$ |
|---|---|---|---|---|---|---|---|---|---|
| | $H_{4\langle110\rangle}$ (Oe) | $H_{2[110]}$ (Oe) | $H_{2[100]}$ (Oe) | $\varepsilon$ (Oe) | $H_{4\langle110\rangle}$ (Oe) | $H_{2[110]}$ (Oe) | $H_{2[100]}$ (Oe) | $\varepsilon$ (Oe) | |
| 0.015 | 250 | -130 | 65 | 330 | 300 | -50 | 20 | 170 | 0.13 |
| 0.05 | 420 | -160 | 80 | 330 | 300 | -50 | 20 | 170 | 0.11 |
| 0.1 | 480 | -40 | 50 | 330 | 300 | -50 | 20 | 170 | 0.08 |
| 0.2 | 350 | -20 | 20 | 330 | 300 | -50 | 20 | 170 | 0.04 |

Supplemental Material

# Ultra-low-power orbital-controlled magnetization switching using a ferromagnetic oxide interface

## I.     Materials and Methods

**1.     Sample growth and characterizations.** The magnetic tunnel junction (MTJ) used in this study was grown on a $TiO_2$-terminated $SrTiO_3$ (STO) (001) substrate via molecular beam epitaxy (MBE) with a shuttered growth technique [Fig. S1(a)]. The fluxes of La, Sr, Mn and Ti were supplied by Knudsen cells. The STO and $La_{0.67}Sr_{0.33}MnO_3$ (LSMO) layers were grown at 730°C with a background pressure of $7\times10^{-4}$ Pa due to a mixture of oxygen (80%) and ozone (20%). The *in situ* reflection high-energy electron diffraction (RHEED) patterns in the [100] direction of the STO and LSMO layers were streaky and bright, indicating good growth conditions and high crystal quality [Fig. S1(b)]. Atomic force microscopy measurements reveal flat terraces and atomic steps with a height of ~ 0.35 nm, which is nearly equal to one unit cell (u.c.) of LSMO (0.39 nm) [Fig. S1(d)]. The magnetic properties of the LSMO layers in the MTJ device were characterized by superconducting quantum interference device (SQUID) magnetometry. The saturation magnetic moment per Mn atom is 3.5 $\mu_B$, close to the value of 3.7 $\mu_B$ for bulk LSMO [Fig. S1(d)].

**2.    Device preparation.** We first deposited a 50-nm-thick Au layer on top of the MTJ sample. A circular mesa measuring 800 μm in diameter was then formed by etching the sample to the half of the bottom LSMO layer, using standard photolithography and Ar ion milling. Another Au contact electrode was formed by depositing a 50-nm-thick Au layer on the etched surface of the bottom LSMO layer. The polarity of the bias voltage $V$ was defined such that the current $I$ flows from the top to the bottom of the mesa under a positive bias. The d$I$/d$V$-$V$ characteristics were numerically obtained from $I$-$V$ data in intervals of 20 mV. The d$I$/d$V$-$V$ curve is nearly symmetric with $V$, which is consistent with the symmetric structure of this LSMO/STO/LSMO MTJ. All transport measurements were carried out at 3.5 K.

**3.    Measurements of the in-plane magnetic-field-angle $\theta_H$ dependence of tunneling magnetoresistance (TMR).** To measure the $\theta_H$ dependence of the TMR, at first we applied a strong magnetic field of 1 T in the direction opposite to $\theta_H$ to align the magnetization directions; we then decreased the magnetic field to zero before the measurements. Subsequently, we started to measure the tunnel resistance $R$ while increasing the magnetic field from zero in the $\theta_H$ direction. The measurements were performed at every 10° step of $\theta_H$. The TMR ratio is defined as follows:

$$\text{TMR} = \frac{[R(\mathbf{H}) - R(0)]}{R(0)} = \frac{P^2[1 - \cos(\Delta\theta)]}{1 + P^2 \cos(\Delta\theta)}. \tag{S1}$$

Here $R(\mathbf{H})$ represents the tunnel resistance under $\mathbf{H}$, and $P$ is the spin polarization of the ferromagnetic layers. The TMR ratio depends on the relative angle $\Delta\theta$ of the magnetization directions between the top and bottom LSMO layers, which differs depending on the direction and intensity of $\mathbf{H}$ due to the different magnetic anisotropy (MA) in the top and bottom LSMO layers. Therefore, by measuring the $\theta_H$ dependence of TMR, we can characterize the MA of the ferromagnetic electrodes.

**4.    Measurements of $R_{P200}$, $R_{AP200}$, $R_{P15}$, and $R_{AP15}$ [= tunnel resistances at zero magnetic field in the parallel (P) and antiparallel (AP) configurations at $V$ = 200 mV and 15 mV] and estimation of $\Delta\theta$.** Here, we explain how $R_{P15}$, $R_{AP15}$, $R_{P200}$, and $R_{AP200}$ were derived. At each $V$ (15 mV or 200 mV), at first we applied a strong $\mu_0\mathbf{H}$ of 1 T in the specific direction ([110] or [$\bar{1}$10]) to align the magnetizations of the top and bottom LSMO layers; we then decreased $\mathbf{H}$ to zero, at which we obtained the $R_{P15}$ or $R_{P200}$ values. Subsequently, we gradually increased $\mathbf{H}$ from zero in the opposite direction while monitoring the tunneling resistance $R$. When the magnetic field became equal to the coercive force of one of the LSMO electrodes, the magnetization of the LSMO electrode switched and $R$ suddenly increased; immediately after that switching, we decreased $\mathbf{H}$ to zero again and obtained the $R_{AP15}$ or $R_{AP200}$ values. At this point, particularly in the case of $V$ = 200 mV where the symmetry of the MA is four-fold and the easy magnetization axes are along [110] and [$\bar{1}$10], the magnetization directions are antiparallel (AP).

From the $R_{P200}$ and $R_{AP200}$ values, using Eq. (S1) for $\Delta\theta$ =180°, we have the following relation:

$$P^2 = \frac{a}{a+2} \ .$$  (S2)

Here,

$$a = \frac{R_{AP200} - R_{P200}}{R_{P200}} = \frac{2P^2}{1-P^2}.$$  (S3)

On the other hand, for an arbitrary angle $\Delta\theta$ between the magnetization vectors of the top and bottom LSMO layers and the corresponding resistance $R$ at $V$ = 200 mV of the

MTJ, we have the following relation:

$$\cos(\Delta\theta) = \frac{P^2 - b}{P^2(1+b)} \quad .$$ (S4)

Here,

$$b = \frac{R - R_{P200}}{R_{P200}} = \frac{P^2[1 - \cos(\Delta\theta)]}{1 + P^2 \cos(\Delta\theta)}.$$ (S5)

From Eqs. (S2) and (S4), we deduced

$$\cos(\Delta\theta) = \frac{a - 2b - ab}{a + ab} \quad .$$ (S6)

From Eqs. (S3) and (S5), we deduced the form

$$\cos(\Delta\theta) = \frac{2R_{P200}R_{AP200} - RR_{AP200} - RR_{P200}}{R(R_{AP200} - R_{P200})} \quad .$$ (S7)

Using Eq. (S7) [= Eq. (1) in the main text] and the experimental values of $R_{P200}$ and $R_{AP200}$, we can determine $\Delta\theta$ from $R$ after the bias sequence when **H**//[110].

**5.     Simulation of the in-plane $\theta_H$ dependence of TMR.** We calculated the $\theta_H$ dependence of TMR using the Stoner-Wohlfarth model. The magnetostatic energy $E$ of an LSMO layer, which has a magnetization **M** in the direction of the angle $\theta_M$ with respect to the in-plane [100] direction, is expressed by

$$E = \frac{MH_{4\langle110\rangle}}{8}\sin^2 2\left(\theta_M - \frac{\pi}{4}\right) + \frac{MH_{2[110]}}{2}\sin^2\left(\theta_M - \frac{\pi}{4}\right) + \frac{MH_{2[100]}}{2}\sin^2(\theta_M) - MH\cos(\theta_M - \theta_H).$$ (S8)

Here, $H_{4\langle110\rangle}$, $H_{2[110]}$, and $H_{2[100]}$ are the anisotropy fields of the biaxial MA along the <110> axes and two uniaxial MA along the [110] and [100] axes, respectively. In this model, $\theta_M$ is determined such that $E$ takes a local minimum. We allowed magnetization reversal by the nucleation and propagation of the 180° domain wall; $\theta_M$ shifts from one local minimum to another local minimum when the energy difference between these points exceeds a threshold energy $\varepsilon$. We fixed $\varepsilon$ at 330 Oe and 170 Oe for the top and bottom LSMO layers, respectively, which were determined by fitting the calculated coercive fields to the magnetic field values where the resistance switching occurs in the TMR measurements when **H** is applied along the [100] easy axis. As explained in Fig. 1(c), when applying a positive (negative) $V$, carrier depletion occurs at the top (bottom) interface between LSMO and STO. Consequently, the quasi Fermi level $E_F$ of the top (bottom) LSMO at the interface with STO is lowered from the $e_g$ band to the $t_{2g}$ band. Therefore, when $V > 0$ the change in the orbital symmetry from $e_g$ to $t_{2g}$ at the quasi $E_F$ only occurs in the top LSMO. Thus, we simplified the fitting process by keeping the set of fitting parameters of the anisotropy fields in the bottom layer fixed and varying only the parameters in the top layer, as summarized in Table 1 in the main manuscript. The TMR ratio was then calculated from $\Delta\theta$ using equation (S1). The $\theta_H$ dependence of TMR reproduced by the Stoner-Wohlfarth model is represented in Fig. S2, which can well explain the experimental data in Fig. 2(d).

## 6. Bias-dependence of the magnetization-direction dependence of the DOS and the deduction of the symmetry components of the density of states (DOS)

We investigated the magnetization-direction dependence of the DOS by measuring tunneling anisotropic magnetoresistance (TAMR); we measured $dI/dV$ by applying **H** = 10 kOe in various directions of $\theta_H$ with a step of 10° at various $V$. Under this strong **H,** the magnetization directions of both LSMO layers are aligned in the **H** direction. Thus, TMR is zero for all $\theta_H$; however, $dI/dV$ depends on $\theta_H$ because the DOS of the ferromagnetic electrodes depends on the direction of the magnetization (// **H**). At each $V$ and $\theta_H$, we define

$$\Delta\left(\frac{dI}{dV}\right) = \left(\frac{dI}{dV} - \langle\frac{dI}{dV}\rangle_{\theta_H}\right) \Big/ \langle\frac{dI}{dV}\rangle_{\theta_H} \times 100 \ \ (\%), \tag{S9}$$

where $\langle\frac{\mathrm{d}I}{\mathrm{d}V}\rangle_{\theta_{\mathrm{H}}}$ is the averaged d$I$/d$V$ over $\theta_{\mathrm{H}}$ at fixed $V$.

We deconvoluted the $\theta_{\mathrm{H}}$ dependence of $\Delta$(d$I$/d$V$) into three symmetry components using the following fitting equation at each $V$:

$$\Delta\left(\frac{\mathrm{d}I}{\mathrm{d}V}\right) = C_{4\langle110\rangle}\cos4\left(\theta_H - \frac{\pi}{4}\right) + C_{2[100]}\cos2\theta_H + C_{2[110]}\cos2\left(\theta_H - \frac{\pi}{4}\right). \text{ (S10)}$$

Note that $C_{4\langle110\rangle}$ and $C_{2[100]}$ are usually observed in LSMO films grown on STO (001) substrates. On the other hand, $C_{2[110]}$ has been hardly reported in bulk LSMO and thus can be attributed to the LSMO/STO interfaces. The values of $C_{4\langle110\rangle}$, $C_{2[100]}$, and $C_{2[110]}$ obtained in this study are summarized as functions of $V$ in Fig. S3(b).

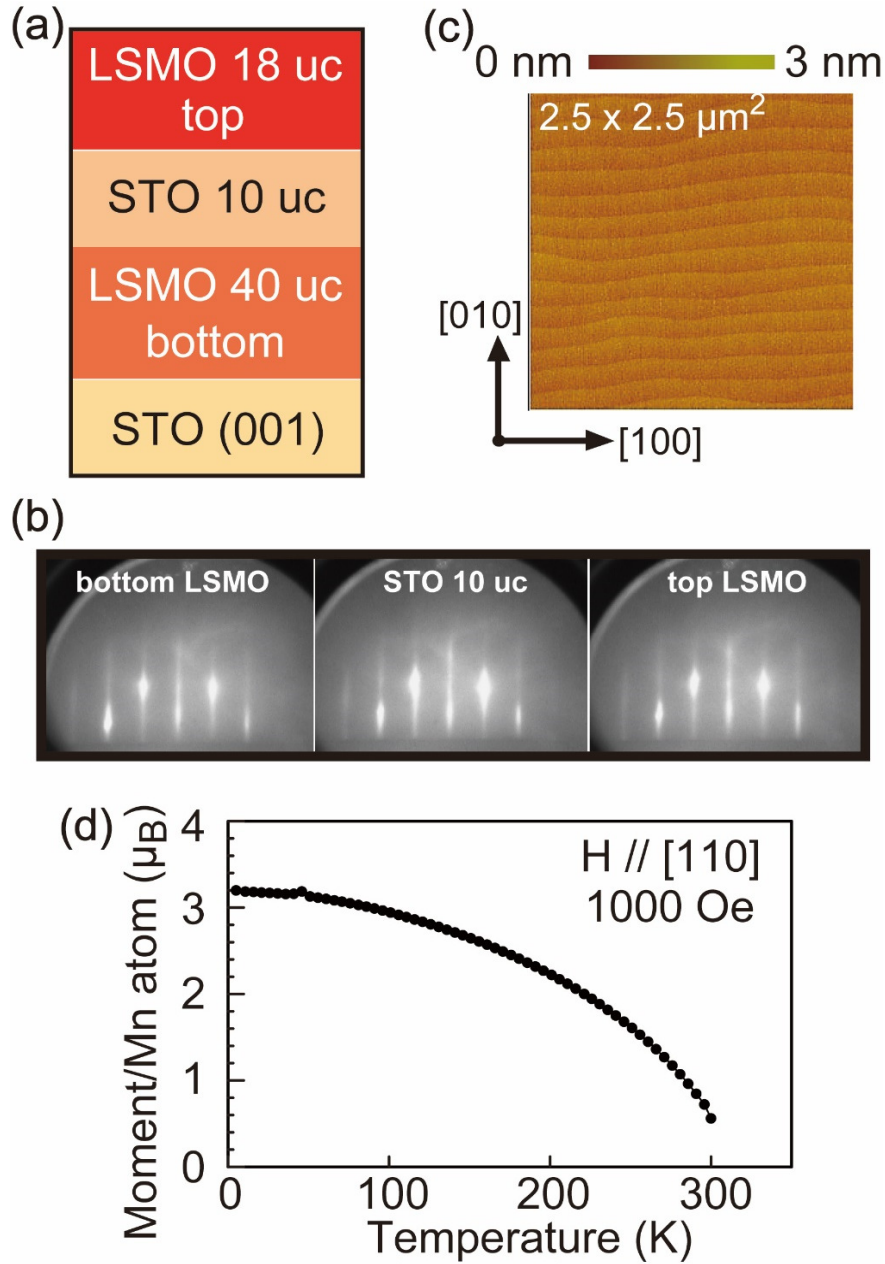

FIG. S1. (a) Schematic cross-sectional structure of the MTJ used in this study. (b) *In situ* RHEED patterns in the [100] direction of the STO and LSMO layers. We observed streaky and bright patterns, indicating good growth conditions and high crystal quality. (c) Atomic force microscopy measurements reveal flat terraces and atomic steps with a height of ~ 0.35 nm, which is nearly equal to one u.c. of LSMO (0.39 nm). (d) Temperature dependence of the magnetic moment per Mn atom, measured with **H** of 1000 Oe applied in the [110] direction.

FIG. S2. Bias-voltage and magnetic-field-direction dependences of tunneling magnetoresistance (TMR) reproduced by the Stoner-Wohlfarth model. Polar colour-mapping plots of the magnetic-field-direction dependence of TMR under a bias voltage $V$ = 15, 50, 100, and 200 mV. The used parameters are summarized in Table 1.

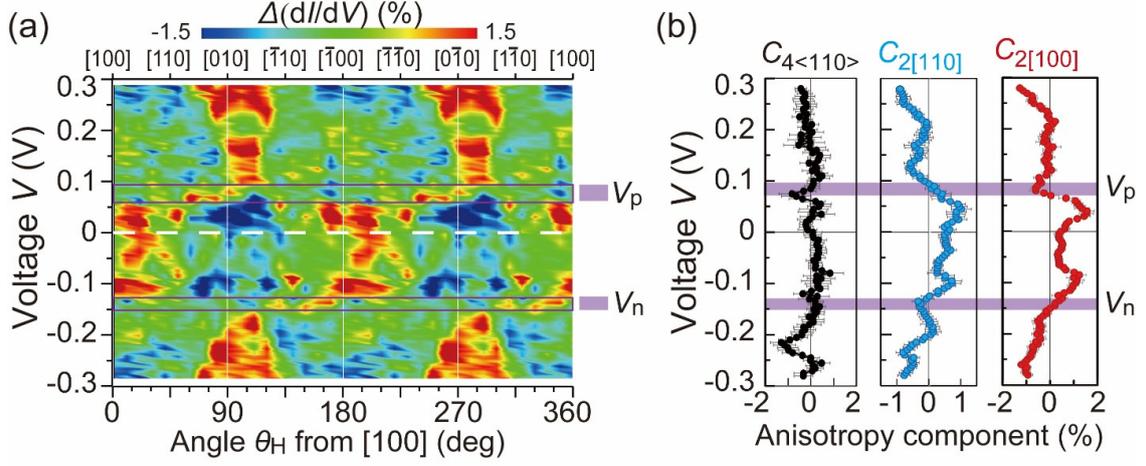

FIG. S3. (a) Color-mapping plot of $\Delta\left(\frac{\mathrm{d}I}{\mathrm{d}V}\right) = \left(\frac{\mathrm{d}I}{\mathrm{d}V} - \langle\frac{\mathrm{d}I}{\mathrm{d}V}\rangle_{\theta_H}\right) / \langle\frac{\mathrm{d}I}{\mathrm{d}V}\rangle_{\theta_H} \times 100$ (%) as a function of $\theta_H$ and $V$. Here, $\theta_H$ is the magnetic-field angle from the [100] axis in the counter-clockwise direction in the film plane, and $\langle\frac{\mathrm{d}I}{\mathrm{d}V}\rangle_{\theta_H}$ is defined as the averaged $\mathrm{d}I/\mathrm{d}V$ over $\theta_H$ at fixed $V$. The direction $\theta_{\mathrm{DOS\text{-}MAX}}$, along which $\mathrm{d}I/\mathrm{d}V$ reaches its maximum, rotates by 90° at $V = V_p$ (= 0.6 − 0.95 V) and $V_n$ [= (− 0.15) − (− 0.13) V ] (purple bands). (b) $V$ dependence of the symmetry components of the DOS, $C_{4\langle110\rangle}$, $C_{2[100]}$, and $C_{2[110]}$. The error bars are the standard errors in the fitting process.

24